\documentclass[sigconf]{acmart}

\usepackage{caption}
\usepackage{subcaption}

\AtBeginDocument{%
  \providecommand\BibTeX{{%
    \normalfont B\kern-0.5em{\scshape i\kern-0.25em b}\kern-0.8em\TeX}}}

\setcopyright{rightsretained} 
\copyrightyear{2020} 
\acmYear{2020} 
\acmConference{SA '20 Posters}{December 04-13, 2020}{Virtual Event, Republic of Korea}
\acmBooktitle{SIGGRAPH Asia 2020 (SA '20 Posters), December 04-13, 2020}
\acmDOI{10.1145/3415264.3425464}
\acmISBN{978-1-4503-8113-0/20/11}

\acmSubmissionID{pos\_164}

\begin{document}

\title{Enhanced Direct Delta Mush}

\author{Serguei Kalentchouk}
\email{skalentchouk@apple.com}
\affiliation{%
  \institution{}
}

\author{Michael Hutchinson}
\email{michael_hutchinson@apple.com}
\affiliation{%
  \institution{}
}

\author{Deepak Tolani}
\email{dtolani@apple.com}

\renewcommand{\shortauthors}{Kalentchouk, Hutchinson and Tolani}

\ccsdesc[500]{Computing methodologies~Animation}

\keywords{skinning, deformation, character animation, delta mush}

\begin{teaserfigure}
  \centering
  \begin{subfigure}[b]{0.3\textwidth}
    \centering
    \includegraphics[width=\textwidth]{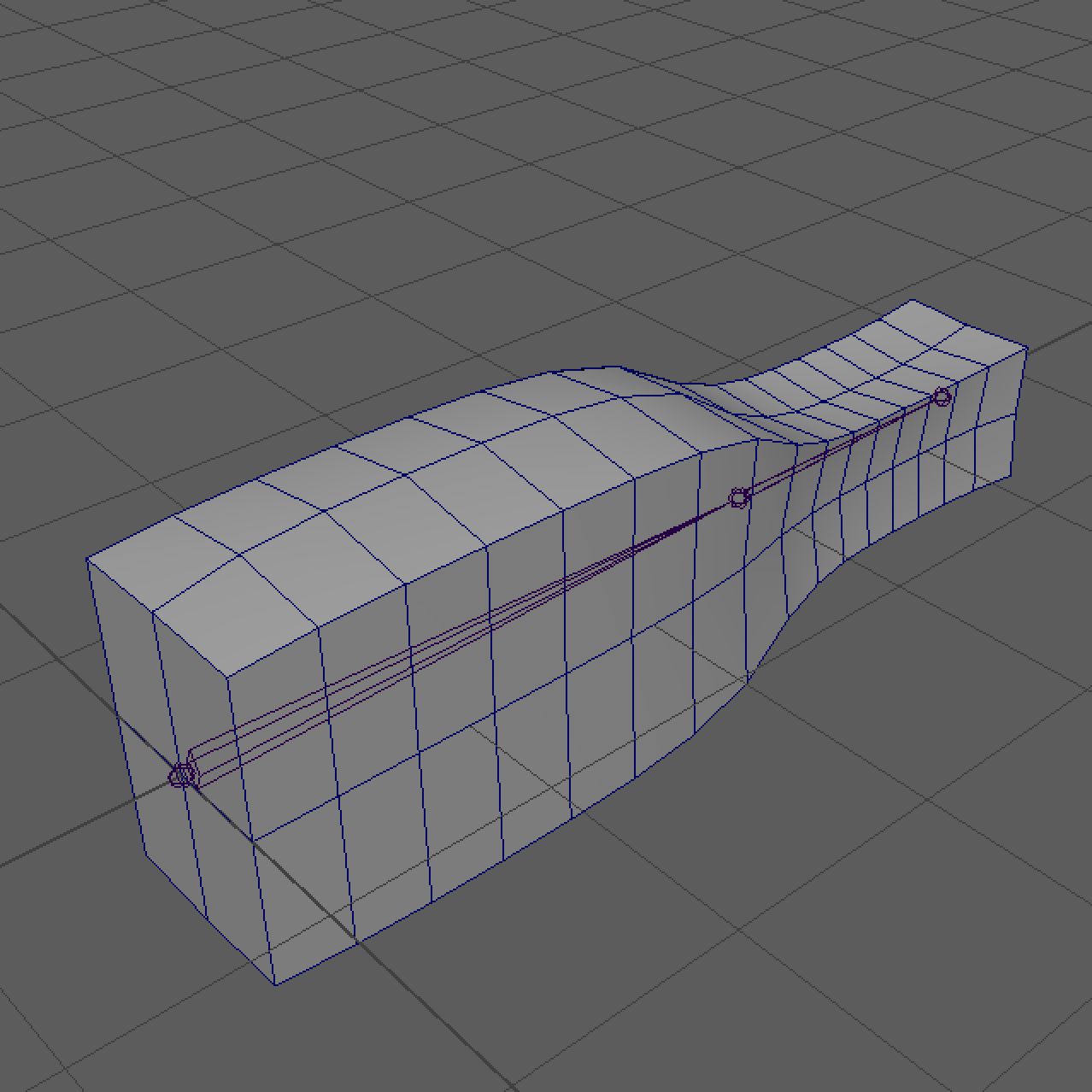}
    \caption{DDM}
    \label{fig:ddm}
  \end{subfigure}
  \hfill
  \begin{subfigure}[b]{0.3\textwidth}
    \centering
    \includegraphics[width=\textwidth]{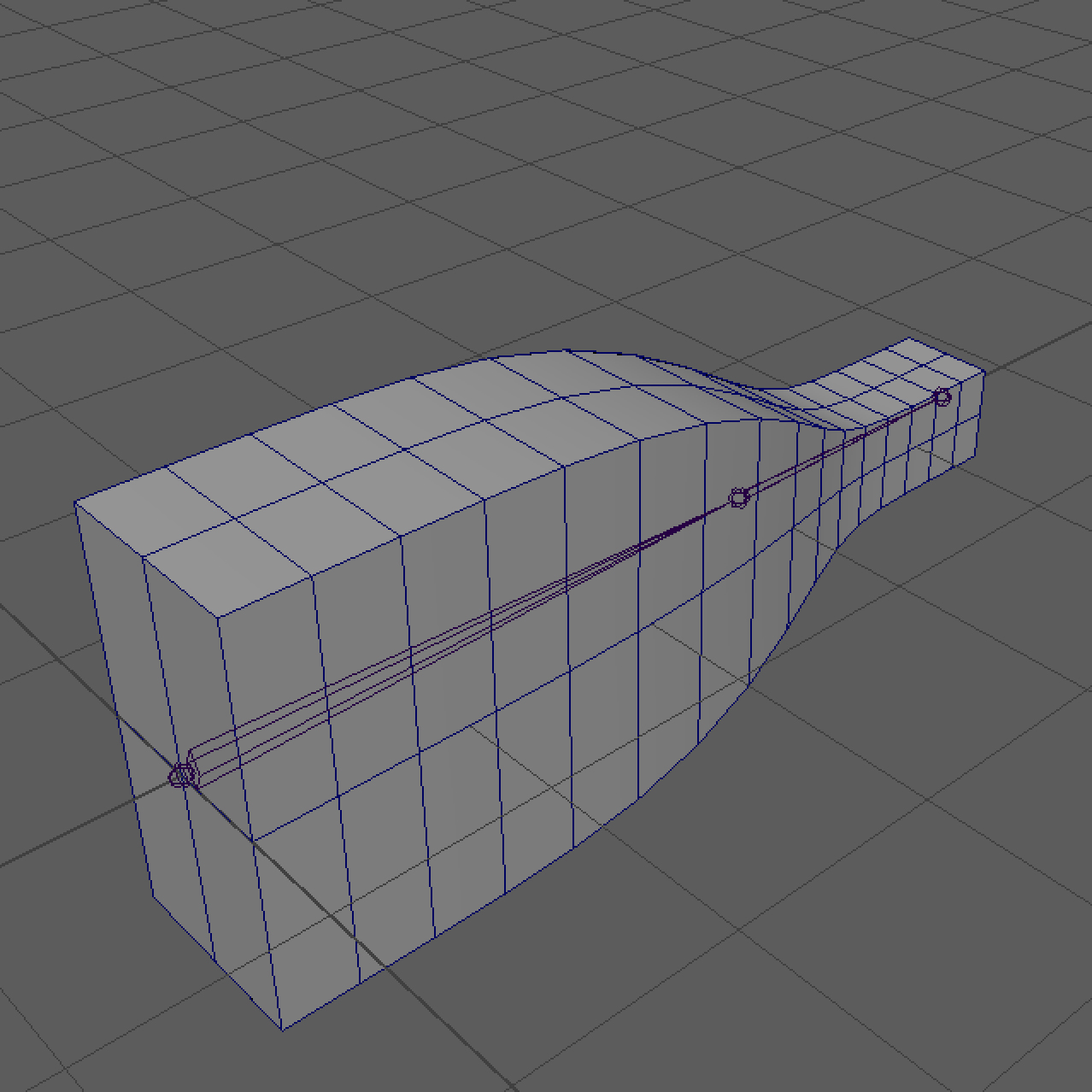}
    \caption{LBS}
    \label{fig:lbs}
  \end{subfigure}
  \hfill
  \begin{subfigure}[b]{0.3\textwidth}
    \centering
    \includegraphics[width=\textwidth]{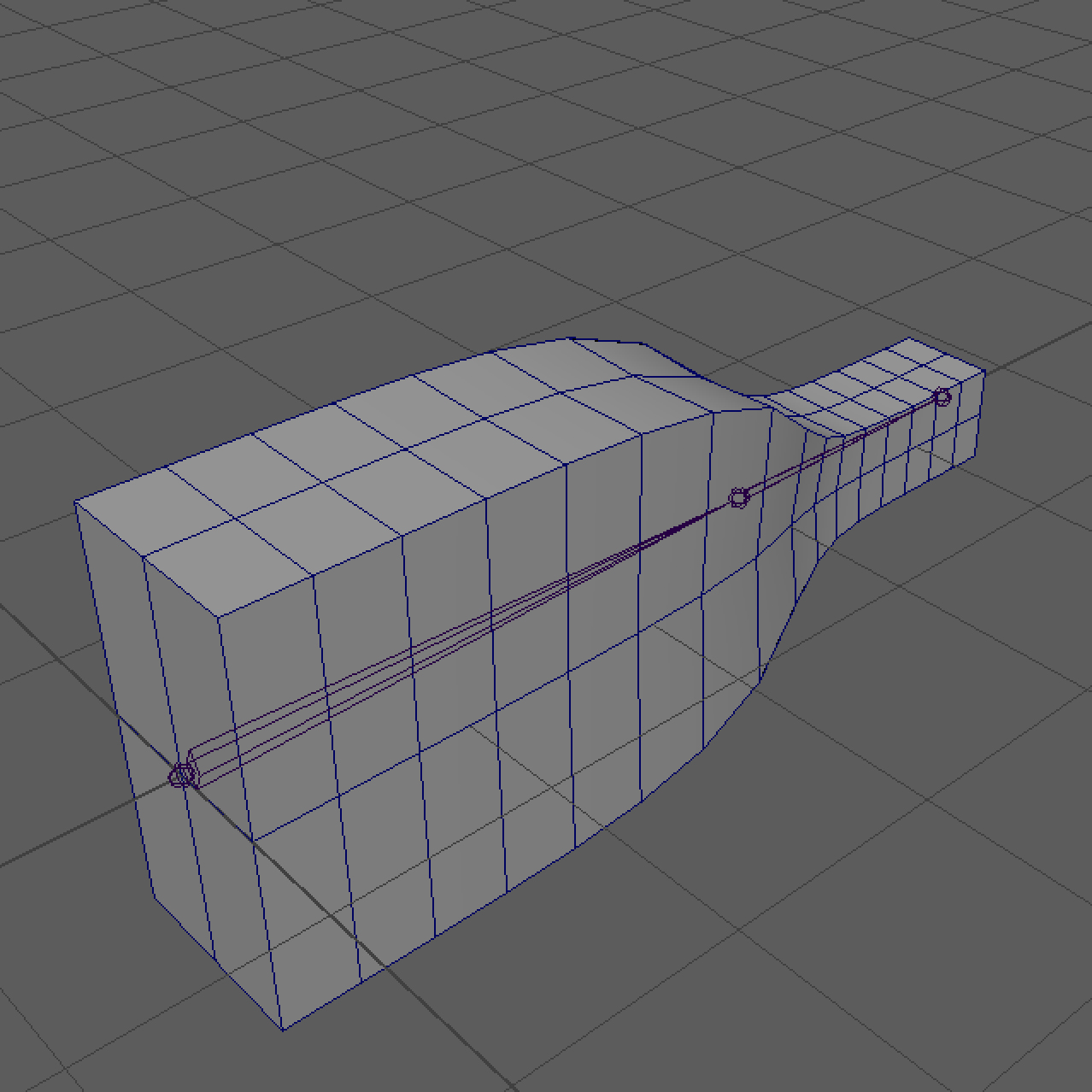}
    \caption{Enhanced DDM}
    \label{fig:eddm}
  \end{subfigure}
  \caption{Skinning deformation under non-rigid transformations, with first joint scaled by 2 in the Y axis and the second joint scaled uniformly by 0.5. DDM (a) distorts the original shape and does not capture the full magnitude of the scaling applied to the vertices.}
  \label{fig:deformation_variants}
\end{teaserfigure}

\maketitle

\section{Introduction}
Direct Delta Mush is a novel skinning deformation technique introduced by Le and Lewis (2019). It generalizes the iterative Delta Mush algorithm of Mancewicz et al (2014), providing a direct solution with improved efficiency and control. Compared to Linear Blend Skinning, Direct Delta Mush offers better quality of deformations and ease of authoring at comparable performance. However, Direct Delta Mush does not handle non-rigid joint transformations correctly which limits its application for most production environments.

This paper presents an extension to Direct Delta Mush that integrates the non-rigid part of joint transformations into the algorithm. In addition, the paper also describes practical considerations for computing the orthogonal component of the transformation and stability issues observed during the implementation and testing.

\begin{figure}[b]
  \centering
  \begin{subfigure}{0.49\linewidth}
    \centering
    \includegraphics[width=\linewidth]{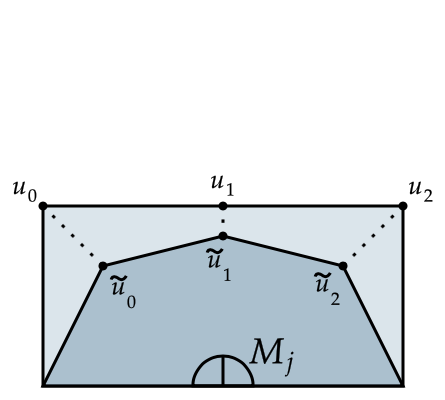}
    \caption{at rest}
  \end{subfigure}
  \hfill
  \begin{subfigure}{0.49\linewidth}
    \centering
    \includegraphics[width=\linewidth]{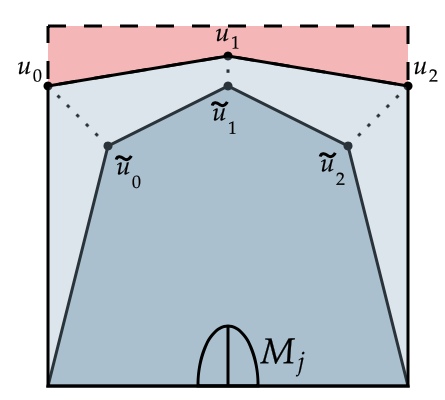}
    \caption{deformed}
  \end{subfigure}
  \caption{Delta mush deformation under non-rigid transformation fails to reconstruct the original shape.}
  \label{fig:dm}
\end{figure}

\section{Non-Rigid Transformations}
“Squash and stretch” is one of the 12 basic principles of animation. In order to accurately model this behavior, it is often necessary to apply non-uniform scale to joint transformations during animation. Unfortunately, Direct Delta Mush (DDM) \cite{Le2019} fails to accurately handle non-rigid joint transformations, as shown in Figure \ref{fig:ddm}, when compared to similar transformations with Linear Blend Skinning (LBS), shown in Figure \ref{fig:lbs}. Not surprisingly, the original Delta Mush (DM) \cite{Mancewicz14} technique exhibits the same artifacts under non-rigid transformations.

Conceptually, DM/DDM applies joint transformations to mesh vertices that have undergone Laplacian smoothing $\mathbf{\tilde{u}}_i\in\mathbb{R}^4$, and then restores the surface detail lost through smoothing by applying the delta between the smoothed and non-smoothed vertices at rest pose. However, while displacements of smoothed and non-smoothed vertices under rigid transformation are equivalent, vertex displacements under non-rigid transformation are not, as seen in Figure \ref{fig:dm}.

The key insight is to compute the displacement of non-smoothed vertices under non-rigid transformation and reintroduce that displacement as a rigid transformation of smoothed vertices.

To that end, the DDM runtime computation is updated as follows. Each joint transformation $\mathbf{M}_j \in \mathbb{R}^{\mathsf{4x4}}$ is factored into a combined scale and shear matrix $ \mathbf{M}_{sj} $ and a rigid only transformation matrix $\mathbf{M}_{rj}$ \cite{Spencer91}. Then, for each vertex $\mathbf{u}_i\in\mathbb{R}^4$ in the rest pose, a displacement under non-rigid transformations is computed as: $\mathbf{d}_{ij}=\mathbf{M}_{sj}\mathbf{u}_i-\mathbf{u}_i$, where $\mathbf{M}_{sj}$ is applied as a local transformation at each joint. Finally, equation 7 rev. of the DDM algorithm is modified to be:
\begin{equation}
\begin{bmatrix}
\mathbf{Q}_i & \mathbf{q}_i \\
\mathbf{p}_i^T & 1
\end{bmatrix} = \displaystyle\sum_{j=1}^m \mathbf{M}_{rj}\mathbf{D}_{ij}\mathbf{\Omega}_{ij}
\end{equation}
where:
\begin{displaymath}
\mathbf{D}_{ij}=\begin{pmatrix}
1 & 0 & 0 & \mathbf{d}_{ij0} \\
0 & 1 & 0 & \mathbf{d}_{ij1} \\
0 & 0 & 1 & \mathbf{d}_{ij2} \\
0 & 0 & 0 & 1
\end{pmatrix}
\end{displaymath}

With this formulation, the non-rigid transformation of each influence joint is then weighted using the precomputed weights, producing the desired deformation seen in Figure \ref{fig:eddm}. This solution supports mixed hierarchies of scale-propagating and scale-compensating joints, which are typical of production rigs in Maya and other digital content creation tools.

\begin{figure}[t]
  \centering
  \begin{subfigure}{0.49\linewidth}
    \centering
    \includegraphics[width=\linewidth]{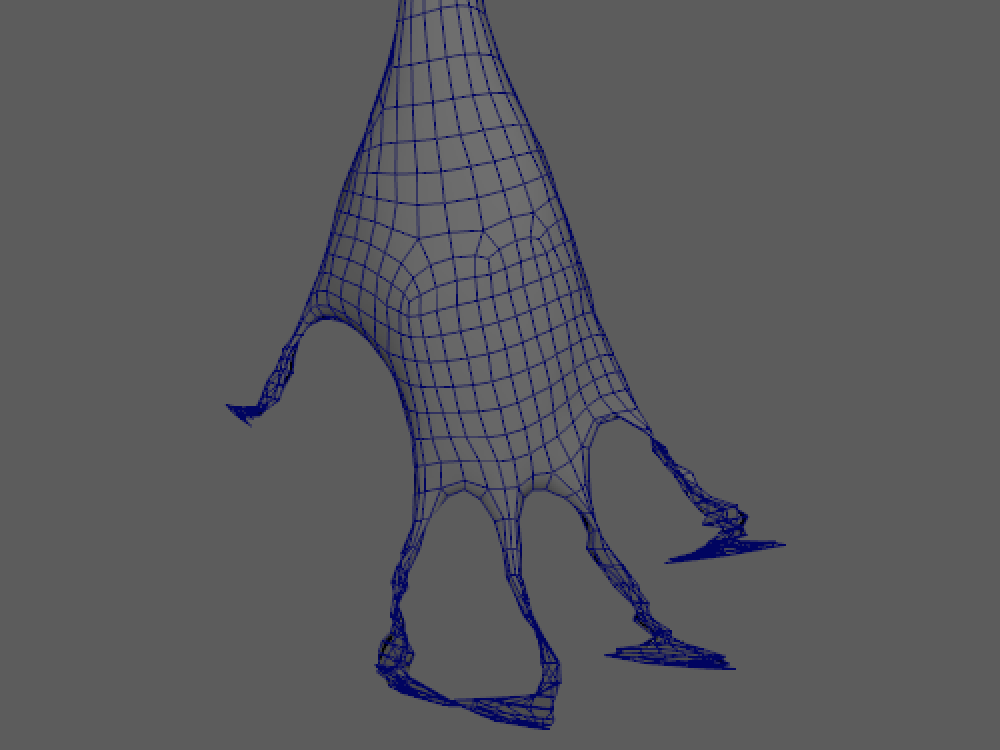}
    \caption{}
    \label{fig:single}
  \end{subfigure}
  \hfill
  \begin{subfigure}{0.49\linewidth}
    \centering
    \includegraphics[width=\linewidth]{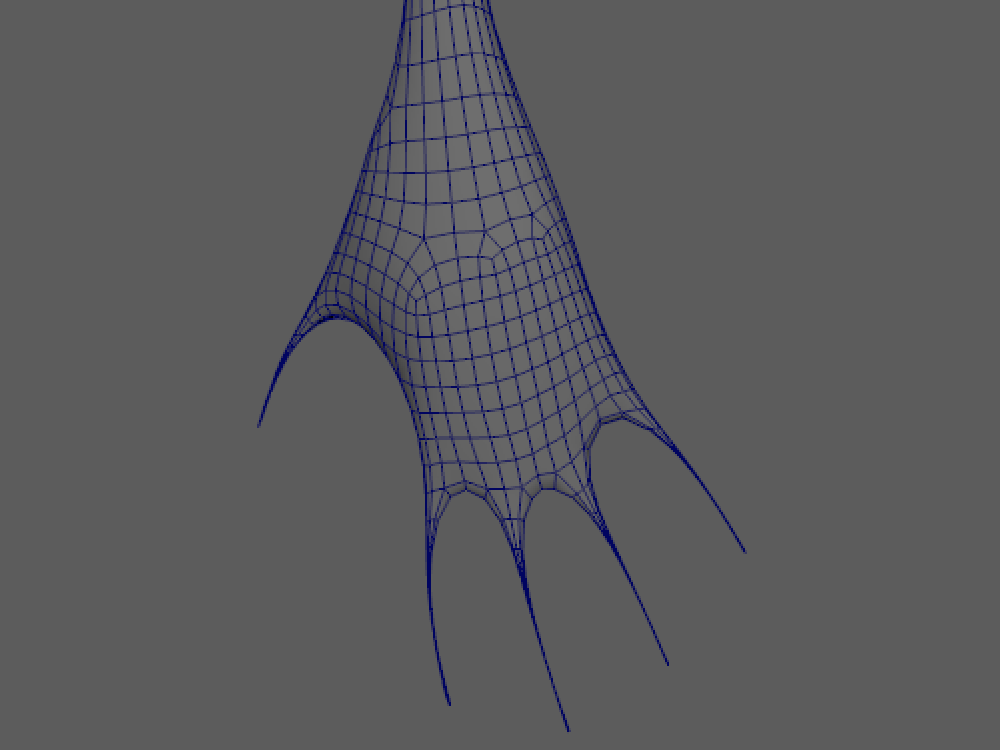}
    \caption{}
    \label{fig:double}
  \end{subfigure}
  \caption{Smoothed vertices transformed by $\mathbf{R}_i$, using a cotangent Laplacian computed with (a) single precision, and (b) with double precision.}
  \label{fig:precision}
\end{figure}

\section{Nearest Orthogonal Matrix}
 To compute the orthogonal component of the vertex transformation, the full model variant of DDM calls for solving the Singular Value Decomposition (SVD) of a $3\mathsf{x}3$ matrix $\mathbf{Q}_i-\mathbf{q}_i\mathbf{p}_i^T=\mathbf{M}_i$, so that $\mathbf{R}_i=\mathbf{U}_i\mathbf{V}_i^T$. Performing a full SVD for each vertex comes at a significant computational cost and poses implementation challenges for real-time environments. Instead, we can take advantage of the fact that $\mathbf{M}$ is a square matrix to consider the Polar Decomposition as a solution to the Procrustes problem. Here, $\mathbf{M}_i=\mathbf{RS}$, where $\mathbf{R}$ is the nearest orthogonal matrix and $\mathbf{S}$ is a symmetric matrix. To solve for $\mathbf{R}$:
\begin{align}
\mathbf{R} = \mathbf{M}_i\mathbf{S}^{-1} \notag\\
\mathbf{M}_i^T\mathbf{M}_i = \mathbf{S}^T\mathbf{R}^T\mathbf{RS}=\mathbf{S}^2\notag\\
\mathbf{R}_i = \mathbf{M}_i(\mathbf{M}_i^T\mathbf{M}_i)^{-1/2}
\end{align}

The square root inverse of the symmetric $3\mathsf{x}3$ matrix can be efficiently computed using a closed form solution, which can be easily implemented as a compute kernel \cite{Franca89}. This formulation provides equivalent results to the full SVD solution at a reduced cost. 

\section{Practical Observations}
While testing with a large range of motion we found instances where the orthogonal matrix $\mathbf{R}_i$ contained reflection as well as rotation. This occurs with both the SVD formulation of Le and Lewis, and our square root formulation. These reflections can result in discontinuous motion and undesirable visual artifacts during animation, making it necessary to restrict the solution to rotation matrices. To do so when using SVD, the formula for $\mathbf{R}_i$ is modified to be \cite{Higham88}:
\begin{equation}
\mathbf{R}_i=\mathbf{U}_i\mathbf{\Sigma'}_i\mathbf{V}_i^T
\end{equation}
where:
\begin{displaymath}
\mathbf{\Sigma'}_i=\begin{pmatrix}
1 & 0 & 0 \\
0 & 1 & 0 \\
0 & 0 & det(\mathbf{U}_i\mathbf{V}_i^T)
\end{pmatrix}
\end{displaymath}

The same correction can be applied to the square root formulation by negating the smallest eigenvalue when the determinant of $\mathbf{M}$ is negative. Note that this requires computing all three eigenvalues using the Stephenson/Sawyers alternative method described by Franca.

Additional testing on more complex geometry revealed that some smoothed vertices were degenerating under the transformation $\mathbf{R}_i\mathbf{p}_i$, as seen in Figure \ref{fig:single}. This issue was also present in the unmodified implementation of DDM. In order to achieve a stable solution we found it necessary to use double precision arithmetic in the calculation of Laplacian cotangent weights.

\bibliographystyle{ACM-Reference-Format}
\bibliography{ddm}

\end{document}